\documentclass[runningheads]{llncs}

\usepackage{amssymb}
\usepackage{graphicx}

\usepackage{url}
\newcommand{\keywords}[1]{\par\addvspace\baselineskip
\noindent\keywordname\enspace\ignorespaces#1}

\begin{document}

\mainmatter

\title{NP by means of lifts and shadows}

\titlerunning{NP by means of lifts and shadows}

\author{G\'abor Kun \and Jaroslav Ne\v{s}et\v{r}il}

\authorrunning{Kun, Ne\v{s}et\v{r}il}

\institute{Department of Mathematics, University of Memphis\\
373 Dunn Hall, Memphis, TN 38152;\\
Department of Applied Mathematics (KAM) and \\
Institute of Theoretical Computer Science (ITI),\\
Charles University, Malostransk\'{e} n\'{a}m 22, Praha;\\
E-mail: kungabor@cs.elte.hu\\
E-mail: nesetril@kam.mff.cuni.cz}


\maketitle

\footnote{Part of this work was supported by ITI and DIMATIA of Charles
University Prague under grant 1M0021620808, by OTKA Grant no.
T043671, NK 67867, by NKTH (National Office for Research and Technology,
Hungary), AEOLUS and also by Isaac Newton Institute (INI)
Cambridge.}

\begin{abstract}
We show that every NP problem is polynomially equivalent to a simple
combinatorial problem: the membership problem for a special class of
digraphs. These classes are defined by means of shadows
(projections) and by finitely many forbidden colored (lifted)
subgraphs. Our characterization is motivated by the analysis of
syntactical subclasses with the full computational power of NP,
which were first studied by Feder and Vardi.

Our approach applies to many combinatorial problems and it induces the 
characterization of coloring problems (CSP) defined by means of shadows.
This turns out to be related to homomorphism dualities. We prove that 
a class of digraphs (relational structures) defined by finitely many 
forbidden colored subgraphs (i.e. lifted substructures) is a CSP class if and only 
if all the the forbidden structures are homomorphically equivalent to 
trees. We show a surprising richness of
coloring problems when restricted to most frequent graph classes.
Using results of Ne\v{s}et\v{r}il and Ossona de Mendez for bounded
expansion classes (which include bounded degree and proper minor
closed classes) we prove that the restriction of every class defined
as the shadow of finitely many colored subgraphs equals to the
restriction of a coloring (CSP) class.

\keywords{digraph, homomorphism, duality, NP, Constraint Satisfaction Problem}

\end{abstract}


\section{Introduction, Background and Previous Work}

Think of $3$-colorability of a graph $G$. This is a well known
hard (and a canonical NP-complete) problem. From the combinatorial
point of view there is a standard way how to approach this problem
(and monotone properties in general): investigate minimal graphs
without this property, denote by $\mathcal F$ the language of all
such critical graphs and define the set $Forb(\mathcal F)$ of all
structures which do not ``contain'' any $F \in \mathcal F$. Then the
language $Forb(\mathcal F)$ coincides with the language of
$3$-colorable graphs. Unfortunately, in the most cases the set
$\mathcal F$ is infinite. However the properties characterized by a
finite set $\mathcal{F}$ are very interesting if we allow {\it
lifts} and {\it shadows}.

Let us  briefly illustrate this by our example of $3$-colorability.
Instead of a graph $G = (V, E)$ we consider the graph $G$ together
with three unary relations $C_1, C_2, C_3$ (i.e. colors of vertices)
which $\it cover$ the vertex set $V$; this structure will be denoted
by $G'$ and called a {\it lift} of $G$ (thus $G'$ has one binary and
three unary relations). There are $3$ {\it forbidden substructures}:
For each $i = 1,2,3$ the single edge graph $K_2$ together with cover
$C_i = \{1,2\}$ and $C_j = \emptyset$ for $j \neq i$ form structure
${\bf F}_i'$ (where the signature of ${\bf F}_i'$ contains one
binary and three unary relations). The language of all $3$-colorable
graphs is just the language $\Phi(Forb({\bf F}_1',{\bf F}_2',{\bf
F}_3'))$, where $\Phi$ is the forgetful functor which transforms
$G'$ to $G$. We call $G$ the {\it shadow} of $G'$.

Clearly this situation can be generalized and one of the main
results of this paper is Theorem \ref{NP} which states that every NP
problem is polynomially equivalent to the membership problem for a
class $\Phi(Forb(\mathcal{F'}))$. Here $\mathcal{F'}$ is a finite
set of (vertex pair)-colored digraphs, $Forb(\mathcal{F'})$ is the
class of all lifted graphs $G'$ for which there is no homomorphism
$F' \longrightarrow G'$ for an $F' \in \mathcal{F'}$. Thus
$Forb(\mathcal{F'})$ is the class of all graphs $G'$ with {\it
forbidden} homomorphisms from $\mathcal{F'}$. (See Section 2 for
definitions.) Theorems \ref{inj} and \ref{full} provide similar
results for forbidden colored subgraphs and for forbidden induced
subgraphs (in both cases vertex colorings suffice).

We should add one more remark. We of course do not only claim that
every problem in NP can be polynomially {\it reduced} to a problem in any
of these classes. This would only mean that each of these classes
contains an NP-complete problem. What we claim is that these classes
have the {\it computational power} of the whole NP class. More
precisely, to each language $L$ in NP there exists a language $M$ in
any of these three classes such that $M$ is {\it polynomially
equivalent} to $L$, i.e. there exist {\it polynomial reductions} of
$L$ to $M$ and $M$ to $L$. E.g. assuming P $\neq$ NP there is a
language in any of these classes that is neither in P nor
NP-complete, since there is such a language in NP by Ladner's
celebrated result \cite{Ladner}.

The expressive power of classes $\Phi(Forb(\mathcal{F'}))$
corresponds to many combinatorially studied problems and presents a
combinatorial counterpart to the celebrated result of Fagin
\cite{FA} who expressed every NP problem in logical terms by means
of an Existential Second Order formula.

The fact that the membership problem for classes
$\Phi(Forb(\mathcal{F'}))$ and their injective and full variants
$\Phi(Forb_{inj}(\mathcal{F'}))$ and
$\Phi(Forb_{full}(\mathcal{F'}))$ have full computational power is
pleasing from the combinatorial point of view as these classes cover
well known examples of hard combinatorial problems: Ramsey type
problems (where as in Theorem \ref{NP} we consider edge colored
graphs), colorings of bounded degree graphs (defined by an
injectivity condition as in Theorem \ref{inj}) and  structural
partitions (studied e.g. in \cite{HELL} as in Theorem \ref{full}).
It follows that, in the full generality, one cannot expect
dichotomies here. On the other side of the spectrum, Feder and Vardi
have formulated the celebrated {\it Dichotomy conjecture} for all
coloring problems (CSP). 

Our main result is Theorem \ref{main}: we give an 
easy characterization of those languages $\Phi(Forb(\mathcal{F'}))$ which
are coloring problems (CSP). This can be viewed as an extension of
the duality characterization theorem for structures \cite{FNT}. We
demonstrate the power of this theorem while reproving some theorems
about the local chromatic number.
In contrast with this we show that the shadow $\Phi(Forb(\mathcal{F'}))$
of a vertex colored class of digraphs $Forb(\mathcal{F'})$ is always a CSP
language when restricted to a bounded expansion class (this notion generalizes
bounded degree and proper minor closed classes) \cite{NPOM}. Our
main tools are {\it finite dualities} \cite{NT,FNT}, {\it restricted
dualities} \cite{NPOM2}, and the {\it Sparse Incomparability Lemma}
\cite{NR,HN}. The detailed proofs can be found in the full version
of this paper \cite{KNEJC}.

\section{Preliminaries}

We consider finite relational structures although in most of the
paper we only deal with digraphs, i.e. relational structures with
just one binary relation.   This itself is one of the main features
of this note: oriented graphs suffice. Digraphs will be denoted by
${\bf A}, {\bf B},\ldots$ (as we want to stress that they may be
replaced by more general structures).

Let $\Gamma$ denote a finite set we refer to as colors. A
$\Gamma$-{\it colored graph (structure)} is a graph (or structure)
together with either a coloring of its vertices or a coloring of all
pairs of vertices by colors from $\Gamma$. Only in Theorem \ref{NP}
we shall consider coloring of all pairs (but in Theorem \ref{NP}
this will play an important role). Thus in the whole paper we shall
undestand by a colored graph a graph with colored vertices. We
denote colored digraphs (relational structures) by $\bf A', \bf B'$
etc. Following the more general notions in category theory we call
$\bf A'$ a {\it lift} of $\bf A$ and $\bf A$ is called the {\it
shadow} of ${\bf A'}$. Thus (vertex-) colored digraphs (structures)
can be also described as {\it monadic} lifts. A homomorphism of
digraphs (relational structures) preserves all the edges (arcs). A
homomorphism of colored digraphs (relational structures) preserves
the color of vertices (pairs of vertices), too. The {\it Constraint
Satisfaction Problem} corresponding to the graph (relational
structure) $\bf A$ is the membership problem for the class of all
graphs (structures) defined by $\{ {\bf B}: {\bf B}$ is
homomorphic to ${\bf A} \}$. We call a mapping between two
(colored) digraphs a {\it full homomorphism} if in addition the
preimage of an edge is an edge. Full homomorphisms have very easy
structure, as every full homomorphism which is onto is a retraction.
The other special homomorphisms we will be interested in are {\it
injective} homomorphisms.

Let $\mathcal F'$ be a finite set of colored relational structures
(digraphs). By $Forb(\mathcal F')$ we denote the set of all colored
relational structures (digraphs) ${\bf A}'$ satisfying ${\bf F}'
\not\longrightarrow \bf A'$ for every ${\bf F}' \in \mathcal F'$.
(If we use injective or full homomorphisms this will be denoted by
$Forb_{inj}(\mathcal{F'})$ or $Forb_{full}(\mathcal{F'})$,
respectively.)

Similarly (well, dually), for the finite set of colored relational
structures (digraphs) $\mathcal D'$ we denote by $CSP(\mathcal D')$
the class of all colored digraphs ${\bf A}'$ satisfying ${\bf A}'
\longrightarrow {\bf D}'$ for some ${\bf D}' \in \mathcal D'$. (This
is sometimes denoted  by $\rightarrow \mathcal D$.) Now suppose that
the classes $Forb(\mathcal F')$ and $CSP(\mathcal D')$ are equal.
Then we say that the pair $(\mathcal {F', D'})$ is a {\it finite
duality}. Explicitly, a finite duality means that the following
equivalence holds for every (colored) relational structure
(digraph):

$$
\forall {\bf F}' \in \mathcal F'
\hspace{3mm}
{\bf F}' \not\longrightarrow {\bf A}' \iff
\exists {\bf D}' \in \mathcal D'
\hspace{3mm}
{\bf A}'  \longrightarrow {\bf D}'.
$$

We say that the structure $\bf A$ is  $\it core$ if every
homomorphism ${\bf A} \longrightarrow {\bf A}$ is an automorphism.
Every finite structure $\bf A$ contains (up to an isomorphism) a
uniquely determined core substructure homomorphically equivalent to
$\bf A$, see \cite{NT} \cite{HN}. The following result was recently proved
in \cite{FNT} and  \cite{NT}. It characterizes finite dualities of
digraphs (or more generally relational structures with a given
signature).

\begin{theorem}\label{FNT} 
For every finite set $\mathcal F$ of (relational)
forests there exists (up to homomorphism equivalence) a finite uniquely
determined set $\mathcal D$ of structures such that $(\mathcal {F,
D})$ forms a finite duality, i.e. $Forb(\mathcal F)=CSP(\mathcal
D)$. Up to homomorphism equivalence there are no other finite
dualities.
\end{theorem}

Let $\Phi$ denote the forgetful functor which corresponds to a
$\Gamma$-colored relational structure (digraph) the uncolored one,
i.e. it forgets about the coloring. We will investigate classes of
the form $\Phi(Forb(\mathcal{F'}))$. We call the pair
$(\mathcal{F'},\mathcal{D})$ {\it shadow duality} if
$\Phi(Forb(\mathcal{F'}))=CSP(\mathcal{D})$. An example of shadow
duality is the language of 3-colorable graphs discussed in the
introduction (or, as can be seen easily, any CSP problem in
general). Finite dualities became much more abundant when we demand
the validity of the above formula just for all graphs from a given
class $\mathcal K$. In such a case we speak about $\mathcal K$-{\it
restricted duality}. It has been proved in \cite{NPOM2} that so
called {\it Bounded Expansion} classes (which include both proper
minor closed classes and classes of graphs with bounded degree) have
a restricted duality for every choice of $\mathcal{F'}$.

The study of homomorphism properties of structures not containing
short cycles (i.e. with a large girth) is a combinatorial problem
studied intensively. The following result has  proved particularly
useful in various applications. It is often called the {\it Sparse
Incomparability Lemma}:

\begin{theorem}
\label{sparse}

Let $k, \ell$ be positive integers and let $\bf A$ be a structure.
Then there exists a structure $\bf B$ with the following properties:

\begin{enumerate}
\item There exists a homomorphism $f: {\bf B} \longrightarrow {\bf A}$;

\item For every structure $\bf C$ with at most $k$ points the following holds:
there exists a homomorphism $ {\bf A} \longrightarrow {\bf C}$ if
and only if there exists a homomorphism $ {\bf B} \longrightarrow
{\bf C}$;

\item $\bf B$ has girth $\geq \ell$.
\end{enumerate}
\end{theorem}


\vspace{6mm}

This result was proved by probabilistic method in \cite{NR} \cite{NZ},
see also \cite{HN}. The polynomial time construction of ${\bf B}$ is
possible, too: in the case of binary relations (digraphs) this was
done in \cite{MN} and for relational structures in
\cite{CSPvsMMSNP}.

\section{Statement of Results}
\subsection{NP by means of finitely many forbidden lifts}

The class SNP consists of all problems expressible by an existential
second-order formula with a universal first-order part \cite{FA}.
The class SNP is computationally equivalent to NP. Feder and Vardi
\cite{FV} have proved that three syntactically defined subclasses of
the class SNP still have the full computational power of the class
NP. We reformulate this result to our combinatorial setting of lifts
and shadows.


\begin{theorem}~\label{NP}
For every language $L \in NP$ there exist a finite set of colors
$\Gamma$ and a finite set of $\Gamma$-colored digraphs
$\mathcal{F'}$, where we color all pairs of vertices such that $L$
is computationally equivalent to the membership problem for
$\Phi(Forb(\mathcal{F'}))$.
\end{theorem}


\begin{theorem}~\label{inj}
For every language $L \in NP$ there exist a finite set of colors
$\Gamma$ and a finite set of $\Gamma$-colored digraphs
$\mathcal{F'}$, (where we color the vertices) such that $L$ is
computationally equivalent to the membership problem for
$\Phi(Forb_{inj}(\mathcal{F'}))$.
\end{theorem}


\begin{theorem}~\label{full}
For every language $L \in NP$ there exist a finite set of colors
$\Gamma$ and a finite set of $\Gamma$-colored digraphs
$\mathcal{F'}$, (where we color the vertices) such that $L$ is
computationally equivalent to the membership problem for
$\Phi(Forb_{full}(\mathcal{F'}))$.
\end{theorem}

\subsection{Lifts and Shadows of Dualities}

It follows from Section $3.1$ that shadows of $Forb$ of a finite set
of colored digraphs, this is classes $\Phi(Forb(\mathcal F'))$,
where $\mathcal{F'}$ is a finite set, have the computational power of
the whole NP. What about finite dualities? Are the shadow dualities
also more frequent? The negative answer is expressed by Theorem
\ref{eqdual2} and shows a remarkable stability of dualities. Towards
this end we first observe that every duality (of lifted structures)
implies a shadow duality:

\begin{theorem}~\label{eqdual1}
Let $\Gamma$ be a finite set of colors and ${\mathcal F'}$ a finite
set of $\Gamma$-colored digraphs (relational structures), where we
color all of the vertices.
Suppose that there exists a finite set of $\Gamma$-colored digraphs
(relational structures) ${\mathcal D'}$ such that $Forb({\mathcal
F'})=CSP({\mathcal D'})$.
Then $\Phi(Forb(\mathcal F')) = CSP(\Phi(\mathcal D'))$.
\end{theorem}

Theorem \ref{eqdual1} may be sometimes reversed: Shadow dualities
may be ``lifted'' in case that lifted graphs have colored vertices
(this is sometimes described as {\it monadic lift}). This is
non-trivial and in fact Theorem \ref{eqdual2} may be seen as the
core of this paper.

\begin{theorem}~\label{eqdual2}
Let $\Gamma$ be a finite set of colors and ${\mathcal F'}$ be a
finite set of $\Gamma$-colored digraphs (relational structures),
where we color all of the vertices. Suppose that $\Phi(Forb(\mathcal
F')) = CSP({\mathcal D})$ for a finite set ${\mathcal D}$ of
digraphs (relational structures). Then
there exists a finite set ${\mathcal D'}$ of 
$\Gamma$-colored digraphs (relational structures) such that
$Forb(\mathcal F') = CSP({\mathcal D'})$.
\end{theorem}

\section{Proofs}

The proofs of Theorems \ref{NP}, \ref{inj} and \ref{full} are in the
full version of this paper \cite{KNEJC}. We do not include them as
they need some new definitions (and space) but nevertheless
basically follow the strategy of \cite{FV}.

Before proving Theorems~\ref{eqdual1} and \ref{eqdual2} we formulate
first a simple lemma which we shall use repeatedly:

\begin{lemma}(lifting)
\label{lifting} Let ${\bf A, B}$ relational structures, homomorphism
$f: {\bf A} \longrightarrow \bf B$, a finite set of colors $\Gamma$
and $\Phi(\bf B') = \bf B$ be given. Then there exists a lift ${\bf
A'}$, such that $\Phi({\bf A'})={\bf A}$ and the mapping $f$ is a
homomorphism ${\bf A'} \longrightarrow \bf B'$ (of colored structures).
\end{lemma}

\begin{proof}(of Theorem \ref{eqdual1})
Suppose that ${\bf A} \in CSP(\Phi(\mathcal D'))$, say ${\bf A} \in
CSP(\Phi({\bf D}'))$. Now for a homomorphism $f: {\bf
A}\longrightarrow \Phi({\bf D}') $ there is at least one lift ${\bf
A}'$ of $\bf A$ such that the mapping $f$ is a homomorphism ${\bf
A}' \rightarrow {\bf D}'$ (here we use Lifting Lemma \ref{lifting}).
Since the pair$({\mathcal F'},{\mathcal D'})$ is a duality ${\bf F}'
\nrightarrow {\bf A'}$ holds for any ${\bf F}' \in \mathcal F'$ and
thus in turn ${\bf A} \in \Phi({\rm Forb}(\mathcal F'))$.

Conversely, let us assume  that ${\bf A'} \in {\rm Forb}(\mathcal
F')$ satisfies $\Phi(\bf A') = \bf A$. But then ${\bf A'} \in
CSP(\mathcal D')$ and thus by the functorial property of $\Phi$ we
have ${\bf A} = \Phi({\bf A'}) \in CSP(\Phi(\mathcal D'))$.
\end{proof}

\begin{proof}(of Theorem \ref{eqdual2})
Assume $\Phi({\rm Forb}(\mathcal F')) = CSP({\mathcal D})$. Our goal
is to find $\mathcal D'$ such that ${\rm Forb}(\mathcal F') =
CSP({\mathcal D'})$. This will follow as a (non-trivial) combination
of Theorems \ref{FNT} and \ref{sparse}. By Theorem \ref{FNT} we know
that if $\mathcal F'$ is a set of (relational) forests then the set
$\mathcal F'$ has a dual set $\mathcal D'$ (in the class of covering
colored structures; we just list all covering colored substructures
of the dual set guaranteed by Theorem \ref{FNT}). It is
$\Phi({\mathcal D}') = \mathcal{D}$ by Theorem \ref{eqdual1}. So
assume to the contrary that one of the structures, say ${\bf F}_0'$,
fails to be a forest (i.e. we assume that one of the components of
${\bf F}_0'$ has a cycle). We proceed by a refined induction (which
will allow us to use more properties of ${\bf F}_0'$) to show that
$\mathcal D'$ does not exist. Let us introduce carefully the
setting of the induction.

We assume shadow duality $\Phi({\rm Forb}(\mathcal F')) =
CSP({\mathcal D})$. Let ${\mathcal D}$ be fixed throughout the
proof. Clearly many sets $\mathcal F'$ will do the job and we select
the set $\mathcal F'$ such that $\mathcal F'$ consists of cores of
all homomorphic images (explicitly: we close $\mathcal F'$ under
homomorphic images and then take the set of  cores of all these
structures). Among all such sets  $\mathcal F'$ we take a set of
minimal cardinality. It will be again denoted by $\mathcal F'$. We
proceed by induction on the size $|\mathcal F'|$ of $\mathcal F'$.

The set ${\rm Forb}(\mathcal F')$ is clearly determined by the
minimal elements of $\mathcal F'$ (minimal in the homomorphism
order). Thus let us assume that one of these minimal elements, say
${\bf F}_0'$, is not a forest. By the minimality of $\mathcal F'$ we
see that we have a proper inclusion $\Phi({\rm Forb}(\mathcal F'
\setminus \{{\bf F}_0'\})) \supset CSP({\mathcal D})$. Thus there
exists a structure $\bf S$ in the difference. But this in turn means
that there has to be a lift $\bf S'$ of $\bf S$ such that ${\bf
F}_0' \longrightarrow \bf S'$ and ${\bf S} \not\rightarrow {\bf D}$
for every ${\bf D} \in \mathcal D$. In fact not only that: as ${\bf
F}_0'$ is a core, as ${\rm Forb}(\mathcal F')$ is homomorphism
closed  and as $\mathcal F'$ has minimal size we conclude that there
exist $\bf S$ and $\bf S'$ such that any homomorphism ${\bf F}_0'
\longrightarrow {\bf S}'$ is a monomorphism (i.e. one-to-one,
otherwise we could replace ${\bf F}_0'$ by a set of all its
homomorphic images - ${\bf F}_0'$ would not be needed).

Now we apply (the second non-trivial ingredient) Theorem
\ref{sparse} to structure $\bf S$ and an $\ell > |X({\bf F}_0')|$:
we find a structure ${\bf S}_0$ with the following properties: ${\bf
S}_0 \longrightarrow \bf S$,  ${\bf S}_0 \longrightarrow {\bf D}$ if
and only if ${\bf S} \longrightarrow {\bf D}$ for every ${\bf D}
\in  {\mathcal D}$ and ${\bf S}_0$ contains no cycles of length
$\leq \ell$. It follows that  ${\bf S}_0 \not\in CSP({\mathcal D})$.
Next we apply Lemma \ref{lifting} to obtain a structure ${\bf S}_0'$
with ${\bf S}_0' \longrightarrow \bf S'$. Now we use that ${\bf
S}_0'$ is a monadic lift and so does not contain cycles of length
$\leq \ell$. Now for any ${\bf F}' \in \mathcal F'$, ${\bf F}' \neq
{\bf F}_0'$ we have ${\bf F}' \nrightarrow {\bf S}_0'$ as ${\bf
S}_0' \rightarrow \bf S'$ and ${\bf F}' \nrightarrow {\bf S}'$. As
the only homomorphism ${\bf F}_0' \longrightarrow {\bf S}'$ is a
monomorphism the only (hypothetical) homomorphism ${\bf F}_0'
\longrightarrow \bf S'$ is also monomorphism. But this is a
contradiction as ${\bf F}_0'$ contains a cycle while ${\bf S}_0'$
has no cycles of length $\leq \ell$. This completes the proof.
\end{proof}

\section{Applications}

\subsection{Classes with bounded expansion}

We study the restriction of classes $\Phi(Forb(\mathcal{F'}))$ to a
class of digraphs with bounded expansion recently introduced in
\cite{NPOM}. These classes are a generalization of proper minor
closed and bounded degree classes of graphs. Using the decomposition
technique of \cite{NPOM} \cite{NPOM2} we can prove that any class
$\Phi(Forb(\mathcal{F'}))$ (for a finite set $\mathcal{F'}$ of
monadic lifts) when restricted to a bounded expansion class equals
to a CSP class (when restricted to the same class).

\begin{theorem}~\label{bexp} Consider the finite set of colors $\Gamma$ and
the class $\Phi(Forb(\mathcal{F'}))$ for a finite set ${\mathcal F'}$ of
$\Gamma$-colored digraphs. Let $\mathcal{C}$ be a class of digraphs
of bounded expansion. Then there is a finite set of digraphs
$\mathcal{D}$ such that $\Phi(Forb(\mathcal{F'})) \cap \mathcal{C} =
CSP(\mathcal{D}) \cap \mathcal{C}$.
\end{theorem}

Consider a monotone, first-order definable class of colored digraphs
$\mathcal{C}$ which is closed under homomorphism and disjoint union. By a
combination with recent results of \cite{ADK} we also obtain (perhaps a bit
surprisingly) that the shadow $\mathcal{C}$ is a CSP language of digraphs. It
remains to be seen to which bounded expansion classes (of graphs and structures) 
this result generalizes.

\subsection{The classes MMSNP and FP - a characterization}

We conclude with an application to descriptive theory of complexity
classes. Recall that the class of languages defined by monotone,
monadic formulas without inequality is denoted by MMSNP ({\it
Monotone Monadic Strict Nondeterministic Polynomial}). (Feder and
Vardi proved that the class MMSNP is computationally equivalent to
the class CSP in a random sense \cite{FV}, this was later
derandomized by the first author \cite{CSPvsMMSNP}.) Madeleine
\cite{M} introduced the class FP of languages defined similarly to
our forbidden monadic lifts of structures.

It has been proved in \cite{M} that the classes FP and MMSNP are
equal. In fact the class MMSNP contains exactly the languages
defined by forbidden monadic lifts.

\begin{proposition} A language of relational structures $L$ is in the class
MMSNP if and only if there is a finite set of colors $\Gamma$ and a
finite set of $\Gamma$-colored relational structures $\mathcal{F'}$
such that $L=\Phi(Forb(\mathcal{F'}))$.
\end{proposition}

Madelaine and Stewart \cite{MStuart} gave a long process to decide
whether an FP language is a finite union of CSP languages. We use
Theorems~\ref{eqdual1} and \ref{eqdual2} and the description of
dualities for relational structures \cite{FNT} to give a short
characterization of a more general class of languages.

\begin{theorem}~\label{MS}
\label{main} Consider the finite set of colors $\Gamma$ and the
language $\Phi(Forb(\mathcal{F'}))$ for a finite set ${\mathcal F'}$
of $\Gamma$-colored digraphs (relational structures).

If no ${\bf F'} \in \mathcal{F'}$ contains a cycle then there is a
finite set of digraphs (relational structures) $\mathcal{D}$ such
that $\Phi(Forb(\mathcal{F'})) = CSP(\mathcal{D})$. If one of the
lifts ${\bf F'}$ in a minimal subfamily of $\mathcal{F'}$ contains a
cycle in its core then the language $\Phi(Forb(\mathcal{F'}))$ is
not a finite union of CSP languages.
\end{theorem}

\begin{proof}
If no ${\bf F}' \in Forb(\mathcal{F'})$ contains a cycle then the
set  $\mathcal F'$ has a dual $\mathcal D'$
by Theorem \ref{FNT}, and the shadow of this set $\mathcal D'$ gives
the dual set $\mathcal{D}$ of the set $\Phi(Forb(\mathcal{F'}))$ (by
Theorem~\ref{eqdual1}). On the other side if one ${\bf F}' \in
Forb(\mathcal{F'})$ contains a cycle in its core and if
$\mathcal{F'}$ is minimal (i.e. $\bf F'$ is needed) then
$Forb(\mathcal{F'})$ does not have a dual.
The shadow of the language $Forb(\mathcal{F'})$ is the language $L$
and consequently this fails to be a finite union of CSP languages by
Theorem~\ref{eqdual2}.
\end{proof}

Theorem \ref{main} may be interpreted as stability of dualities for
finite structures. While shadows of the classes $Forb(\mathcal{F'})$
are computationally equivalent to the whole NP, the shadow dualities
are not bringing anything new: these are just shadows of dualities.
In other words: the coloring problems in the class MMSNP are just
shadow dualities. This holds for graphs as well for relational
structures.

\subsection{On the local chromatic number}

Now we apply Theorem~\ref{MS} in the analysis of local 
chromatic number introduced in \cite{EFHKR} (see also \cite{ST}):
we say that a graph $G$ is locally
$(a, b)$-colorable if there exists a proper coloring of $G$ by $b$ colors
so that every (closed) neighborhood of a vertex of $G$ gets at most $a$
colors. It follows from \cite{EFHKR} that the class of all locally $(a,
b)$-colorable graphs is of the form CSP($U(a, b)$) for an explicitely constructed
graph $U(a, b)$. We conclude this paper with an indirect proof of this
result with an application to complexity:

\begin{proposition}
Let $a,b$ be integers and consider the membership problem for the class of locally
$(a,b)$-colorable graphs. This is actually a Constraint Satisfaction Problem which 
is NP-complete if $a,b \geq 3$ and it is polynomial time solvable else.
\end{proposition}

\begin{proof}
Consider the color set 
$\Gamma=\{1, \dots ,b \}$ and the following set $\mathcal{F'}$ of $\Gamma$-colored
undirected graphs. Let $\mathcal{F'}$ consist of all monochromatic edges (colored by
any of the $b$ colors) and all the stars with $a+1$ vertices colored by at least $a+1$ 
colors. The corresponding 
language is exactly the required one: a graph $G$ is in the language iff it
admits a proper $\Gamma$-coloring, this is no monochromatic edge is homomorphic
to the colored graph, such that the neighbourhood of every vertex (including the
vertex itself) has at most $a$ different colors, i.e. no star with $a+1$ vertices
of different color is homomorphic to it. Since $\mathcal{F'}$ consists of colored
trees this will be a CSP language by Theorem~\ref{MS}.

Hell and the second author proved that CSP problems defined by undirected graphs are 
in P if the graph is bipartite and NP-complete else \cite{HN}. We do not determine which 
graph defines this particular CSP problem (of locally $(a,b)$-colorable graphs). But if 
$a,b \geq 3$ then we know that it contains the triangle if, so the problem is NP-complete.
It is easy to see that this membership problem is in P if $a<3$ or $b<3$.
\end{proof}

\section{Summary and Future Work}

We found a computationally equivalent formulation of the class NP by
means of finitely many forbidden lifts of very special type. An
ambitious project would be to find an equivalent digraph coloring
problem for a given NP language really effectively (in human sense,
our results provide a polynomial time algorithm). For example it
would be nice to exhibit a vertex coloring problem that is
polynomially equivalent to the graph isomorphism problem. In general
this mainly depends on how to express the problem in terms of logic.
The next class we seem to be able to deal with are coloring problems
of structures with an equivalence relation. Another good candidate
are lifts using linear order. This promises several interesting
applications which were studied earlier in a different setting.

We also proved that shadow dualities and lifted monadic dualities
are in $1-1$ correspondence. This abstract result has several
consequences and streamlines some earlier results  in descriptive complexity theory
(related to MMSNP and
CSP classes). The simplicity of this approach suggests some other
problems. It is tempting to try to relate Ladner's diagonalization
method \cite{Ladner} in this setting (as it was pioneered by
Lov\'asz and G\'acs \cite{GL} for NP$\cap$coNP in a similar
context). The characterization of Lifted Dualities is beyond reach
but particular cases are interesting as they generalize results of
\cite{NT} \cite{FNT} and as the corresponding duals present polynomial
instances of CSP.

But perhaps more importantly, our approach to the complexity subclasses 
of NP is based on lifts and shadows as a combination of algebra, combinatorics 
and logic. We believe that it has further applications and that it forms a useful
paradigm.

\end{document}